\documentclass[pra,10pt,showpacs]{revtex4}
\usepackage{graphicx,amsmath,amssymb}

\newcommand{\beq}{\begin{equation}}
\newcommand{\eeq}{\end{equation}}
\newcommand{\beqa}{\begin{eqnarray}}
\newcommand{\eeqa}{\end{eqnarray}}
\newcommand{\ket}[1]{| #1 \rangle}

\begin{document}

\title{A class of entanglement monotones for general pure multipartite states based on complex projective variety}

\author{Hoshang Heydari}
\email{hoshang@imit.kth.se} \affiliation{Institute of Quantum
Science, Nihon University, 1-8 Kanda-Surugadai, Chiyoda-ku, Tokyo
101- 8308, Japan}

\date{\today}

\begin{abstract}
We construct a measure of entanglement for general pure
multipartite states based on Segre variety. We also construct a
class of entanglement monotones  based on the Pl\"{u}cker coordinate
equations of the Grassmann variety. Moreover, we discuss and compare
these measures of entanglement.
\end{abstract}


\maketitle

\section{Introduction}
A very interesting measure of entanglement for multipartite states
is usually called entanglement monotones, e.g., a measure of
entanglement which is invariant under stochastic local quantum
operation and classical communication(SLOCC)\cite{Dur00}. For
example, F. Verstraete et al. \cite{Vers2} have presented a general
mathematical framework to describe local equivalence classes of
multipartite quantum states under the action of local unitary and
local filtering operations. Their analysis has lead to the
introduction of entanglement measures for the multipartite states,
and the optimal local filtering operations maximizing these
entanglement monotones were obtained.
E. Briand, \cite{Briand2} et al. have  studied  the invariant theory
of trilinear forms over a three-dimensional complex vector space,
and apply it to investigate the behavior of pure entangled
three-partite qutrit states and their normal forms SLOCC operations.
They described the orbit space of the SLOCC group
$SL(3,\mathbf{C})^{\times 3}$ both in its affine and projective
versions in terms of a very symmetric normal form parameterized by
three complex numbers. They have also shown that the structure of
the sets of equivalent normal forms is related to the geometry of
certain regular complex polytopes.
A. Miyake and M. Wadati \cite{Miyake} have  explored quantum search
from the geometric viewpoint of a complex projective space.
Recently, P\'{e}ter L\'{e}vay \cite{Levay1} have constructed a class
of multi-qubit entanglement monotones which was based on
construction of C. Emary \cite{Emary}. His construction is based on
bipartite partitions of the Hilbert space and the invariants are
expressed in terms of the Pl\"{u}cker coordinates of the
Grassmannian.
We have also constructed entanglement monotones for multi-qubit
states based on Pl\"{u}cker coordinate equations of Grassmann
variety, which are central notion in geometric invariant theory
\cite{Hosh4}.  In this paper, we will construct a class of
entanglement monotones  for  multipartite states based on some
complex projective algebraic variety. In particular, in section
\ref{segresec} we will derive an entanglement measure for general
multipartite states based on Segre variety and in section
\ref{grassmann}, we will construct a class of entanglement monotones
for multipartite states, which coincides with the measure of
entanglement  based on the Segre variety by construction.
 Now, let us start by denoting
a general, pure, composite quantum system with $m$ subsystems
$\mathcal{Q}=\mathcal{Q}^{p}_{m}(N_{1},N_{2},\ldots,N_{m})
=\mathcal{Q}_{1}\mathcal{Q}_{2}\cdots\mathcal{Q}_{m}$, consisting of
the pure state $
\ket{\Psi}=\sum^{N_{1}}_{k_{1}=1}\sum^{N_{2}}_{k_{2}=1}\cdots\sum^{N_{m}}_{k_{m}=1}
$ $\alpha_{k_{1},k_{2},\ldots,i_{m}} \ket{k_{1},k_{2},\ldots,k_{m}}
$ and corresponding the Hilbert space $
\mathcal{H}_{\mathcal{Q}}=\mathcal{H}_{\mathcal{Q}_{1}}\otimes
\mathcal{H}_{\mathcal{Q}_{2}}\otimes\cdots\otimes\mathcal{H}_{\mathcal{Q}_{m}}
$, where the dimension of the $j$th Hilbert space is given  by
$N_{j}=\dim(\mathcal{H}_{\mathcal{Q}_{j}})$. We are going to use
this notation throughout this paper. In particular, we denote a pure
two-qubit state by $\mathcal{Q}^{p}_{2}(2,2)$. Next, let
$\rho_{\mathcal{Q}}$ denote a density operator acting on
$\mathcal{H}_{\mathcal{Q}}$. The density operator
$\rho_{\mathcal{Q}}$ is said to be fully separable, which we will
denote by $\rho^{sep}_{\mathcal{Q}}$, with respect to the Hilbert
space decomposition, if it can  be written as $
\rho^{sep}_{\mathcal{Q}}=\sum^\mathrm{N}_{k=1}p_k
\bigotimes^m_{j=1}\rho^k_{\mathcal{Q}_{j}},~\sum^N_{k=1}p_{k}=1 $
 for some positive integer $\mathrm{N}$, where $p_{k}$ are positive real
numbers and $\rho^k_{\mathcal{Q}_{j}}$ denotes a density operator on
Hilbert space $\mathcal{H}_{\mathcal{Q}_{j}}$. If
$\rho^{p}_{\mathcal{Q}}$ represents a pure state, then the quantum
system is fully separable if $\rho^{p}_{\mathcal{Q}}$ can be written
as
$\rho^{sep}_{\mathcal{Q}}=\bigotimes^m_{j=1}\rho_{\mathcal{Q}_{j}}$,
where $\rho_{\mathcal{Q}_{j}}$ is the density operator on
$\mathcal{H}_{\mathcal{Q}_{j}}$. If a state is not separable, then
it is said to be an entangled state. In the next section we will
construct the Segre variety for multipartite states, which is  a
complex projective variety. However, we will not give any
introduction to this complex space. The reader unfamiliar with this
topic could look at the standard references for the complex
projective variety, namely \cite{Griff78,Mum76}.

\section{Segre variety and multipartite entanglement measure } \label{segresec}

In this section, we will define the Segre variety for a
multi-projective space \cite{Hosh5} and then based on this variety,
we will construct an entanglement measure for general pure
multipartite states. For example, we can construct a projective
variety of $\mathbf{CP}^{N_{1}-1}\times\mathbf{CP}^{N_{2}-1}
\times\cdots\times\mathbf{CP}^{N_{m}-1}$ by its Segre embedding. Let
$(\alpha_{1},\alpha_{2},\ldots,\alpha_{N_{i}})$  be points defined
on $\mathbf{CP}^{N_{i}-1}$. Then the Segre map
\begin{equation}
\begin{array}{ccc}
  \mathcal{S}_{N_{1},\ldots,N_{m}}:\mathbf{CP}^{N_{1}-1}\times\mathbf{CP}^{N_{2}-1}
\times\cdots\times\mathbf{CP}^{N_{m}-1}&\longrightarrow&
\mathbf{CP}^{N_{1}N_{2}\cdots N_{m}-1}\\
 ((\alpha_{1},\alpha_{2},\ldots,\alpha_{N_{1}}),\ldots,
 (\alpha_{1},\alpha_{2},\ldots,\alpha_{N_{m}})) & \longmapsto&
 (\ldots,\alpha_{i_{1},i_{2},\ldots, i_{m}},\ldots). \\
\end{array}
\end{equation}
is well defined for $\alpha_{i_{1}i_{2}\cdots i_{m}}$,$1\leq
i_{1}\leq N_{1}, 1\leq i_{2}\leq N_{2},\ldots, 1\leq i_{m}\leq
N_{m}$ as a homogeneous coordinate-function on
$\mathbf{CP}^{N_{1}N_{2}\cdots N_{m}-1}$. Moreover, the image of the
Segre map is a complex projective variety, which is called Segre
variety. Now, let us consider the composite quantum system
$\mathcal{Q}^{p}_{m}(N_{1},N_{2},\ldots,N_{m})$. Then, the Segre
ideal
$\mathcal{I}^{1}_{\mathcal{Q}_{1}\models\mathcal{Q}_{2}\mathcal{Q}_{3}\cdots\mathcal{Q}_{m}}$
representing if a subsystem $\mathcal{Q}_{1}$ that is unentangled
with  $\mathcal{Q}_{2}\mathcal{Q}_{3}\cdots\mathcal{Q}_{m}$ is
generated by the six 2-by-$2$ minors of
\begin{equation}\label{matrix1}
  \mathrm{Mat}^{1}_{N_{1},N_{2}\cdots N_{m}}=\left(%
\begin{array}{cccc}
  \alpha_{1,1,\ldots,1} & \alpha_{1,1,\ldots,2} &  \ldots & \alpha_{1,N_{2},\ldots,N_{m}}\\
  \alpha_{2,1,\ldots,1} & \alpha_{2,1,\ldots,2} & \ldots & \alpha_{2,N_{2},\ldots,N_{m}}\\
  \vdots & \vdots & \ldots & \vdots\\
  \alpha_{N_{1},1,\ldots,1} & \alpha_{N_{1},1,\ldots,N_{m}} & \ldots & \alpha_{N_{1},N_{2},\ldots,N_{m}}\\
\end{array}%
\right),
\end{equation}
The other $m-1$ Segre ideal of multi-projective space, that is,
$\mathcal{I}^{2}_{\mathcal{Q}_{2}\models\mathcal{Q}_{1}\mathcal{Q}_{3}\cdots\mathcal{Q}_{m}},\ldots,
\mathcal{I}^{m}_{\mathcal{Q}_{m}\models\mathcal{Q}_{1}\mathcal{Q}_{2}\cdots\mathcal{Q}_{m-1}}$
can be defined  by permutation indices of elements of matrix
$\mathrm{Mat}^{1}_{N_{1},N_{2}\cdots N_{m}}$. For example,
$\mathcal{I}^{j}_{\mathcal{Q}_{j}\models\mathcal{Q}_{1}\mathcal{Q}_{2}
\cdots\widehat{\mathcal{Q}}_{j}\cdots\mathcal{Q}_{m-1}}$ is
generated by 2-by-2 minors of
$\mathrm{Mat}^{j}_{N_{j},N_{1}N_{2}\cdots \widehat{N}_{j}\cdots
N_{m}}$, where the hat over $j$th-subsystem
$\widehat{\mathcal{Q}}_{j}$ indicate that we should delete this
subsystem from right hand side of the Segre ideal. Now, based on
2-by-2 minors of these ideals we can construct an entanglement
measure for general pure multipartite states. So, let
$\mathcal{M}_{\nu,\mu}(\mathrm{Mat}^{j}_{N_{j},N_{1}N_{2}\cdots
\widehat{N}_{j}\cdots N_{m}})$ denotes the 2-by-2 minors of matrix
$\mathrm{Mat}^{j}_{N_{j},N_{1}N_{2}\cdots \widehat{N}_{j}\cdots
N_{m}}$, which is generated by rows $\nu$ and $\mu$, where
$\nu<\mu<N_{j}$. Then an entanglement measure for a general pure
multipartite states is give by
\begin{eqnarray}\label{segre}
\mathcal{E}(\mathcal{Q}^{p}_{m}(N_{1},\ldots,N_{m}))&=&\left(\mathcal{N}_{m}\sum^{m}_{j=1}\sum_{\forall
\nu>\mu=1}|\mathcal{M}_{\nu,\mu}(\mathrm{Mat}^{j}_{N_{j},N_{1}N_{2}\cdots
\widehat{N}_{j}\cdots N_{m}})|^{2}\right)^{1/2}.
\end{eqnarray}
By construction, that is, the definition of the Segre ideals,  this
measure vanish on separable set of a  general pure multipartite
state. However, we need to show that this measure is an entanglement
monotones. This can be done using the construction of entanglement
monotones based on the Pl\"{u}cker coordinates of the Grassmann
variety, which we will discuss in the following section.

\section{Grassmann variety}\label{grassmann}
 In this section, we will define the
Grassmann variety \cite{Mum94} and then we will construct a measure
of entanglement based on Pl\"{u}cker coordinate equations of
Grassmann variety. Let $\mathrm{Gr}(r,d)$ be the Grassmann variety
of the $r-1$-dimensional linear projective subspaces of
$\mathbf{CP}^{d-1}$. Now, we can embed $\mathrm{Gr}(r,d)$ into
$\mathbf{P}(\bigwedge^{r}(\mathbf{C}^{d}))=\mathbf{CP}^{\mathcal{N}}$,
$\mathcal{N}={\small\left(%
\begin{array}{c}
  d \\
  r \\
\end{array}%
\right)}-1$, by using the Pl\"{u}cker map
$L\longrightarrow\bigwedge^{r}(L)$, where the exterior product
$\bigwedge^{r}(\mathbf{C}^{d})$ for $1\leq r \leq d$ is a subspace
of $\mathbf{C}^{N_{1}}\otimes\cdots\otimes\mathbf{C}^{N_{m}}$,
spanned by the anti-symmetric tensors. The Pl\"{u}cker coordinates
$P_{i_{1},i_{2},\ldots,i_{r}}, 1\leq i_{1}<\cdots<i_{r}\leq d$ are
the projective coordinates in this projective space. Next, let
$\mathbf{C}[\Lambda(r,d)]$ be a polynomial ring with the Pl\"{u}cker
coordinates $P_{J}$ indexed by elements of the set $\Lambda(r,d)$ of
ordered $r$-tuples in $\{1,2,\ldots d\}$ as its variables. Then the
image of the map $\kappa:\mathbf{C}[\Lambda(r,d)]\longrightarrow
\mathrm{Pol}(\mathrm{Mat}_{r,d})$, which assigns
$P_{i_{1}i_{2}\ldots i_{r}}$ the bracket polynomial
$[i_{1},i_{2},\ldots, i_{r}]$ (the bracket function on the
$\mathrm{Mat}_{r,d}$, whose values on a given matrix is equal to the
maximal minor formed by the columns from a set of $\{1,2,\ldots,
d\}$) is equal to the sub-ring of the invariant of the polynomials.
Moreover, the kernel $\mathcal{I}_{r,d}$ of the map $\kappa$ is
equal to the homogeneous ideal of the Grassmann in its Pl\"{u}cker
embedding. Furthermore, the homogeneous ideal $\mathcal{I}_{r,d}$
defining $\mathrm{Gr}(r,d)$ in its Pl\"{u}cker embedding is
generated by the quadratic polynomials
\begin{eqnarray}
\mathcal{P}_{I,J}=\sum^{r+2}_{t=1}(-)^{t}P_{i_{1},\ldots,
i_{r-1},j_{t}}P_{j_{1}\ldots j_{t-1}j_{t+1},\ldots,j_{r+1}},
\end{eqnarray}
where $I=(i_{1}\ldots i_{r-1}), 1\leq i_{1}<\cdots<i_{r-1}< j_{i}$,
and $J=(j_{1},\ldots ,j_{r+1}), 1\leq j_{1}<\cdots<j_{r+1}\leq d$
are two increasing sequences of numbers from the set
$\{1,2,\ldots,d\}$. Note that the equations $\mathcal{P}_{I,J}=0$
define the Grassmannian $\mathrm{Gr}(r,d)$ are called the
Pl\"{u}cker coordinate equations. For example, for
$\mathrm{Gr}(2,d)$ and $r=2$, we have
\begin{eqnarray}
\mathcal{P}_{I,J}&=&\sum^{4}_{t=1}(-)^{t}P_{i_{1},j_{t}}P_{j_{1}\ldots
j_{t-1}j_{t+1},\ldots,j_{3}}\\\nonumber&=&
-P_{i_{1},j_{1}}P_{j_{2},j_{3}}+P_{i_{1},j_{2}}P_{j_{1},j_{3}}-P_{i_{1},j_{3}}P_{j_{1},j_{2}},
\end{eqnarray}
where $I=(i_{1})$, and $J=(j_{1},j_{2},j_{3})$. Note that, by its
construction, the Grassmannian $\mathrm{Gr}(2,d)$ is invariant under
$SL(2,\mathbf{C})$. Now, let us consider a quantum system
$\mathcal{Q}^{p}_{m}(2,2,\ldots,2)$, where in this case we have
$r=N_{j}=2$ and $d=N_{1}N_{2}\cdots \widehat{N}_{j}\cdots
N_{m}=2^{m-1}$. Then, we define
\begin{eqnarray}
\mathcal{E}_{I,J}(\mathrm{Mat}^{j}_{N_{j},N_{1}N_{2}\cdots
\widehat{N}_{j}\cdots N_{m}})
&=&\sum^{4}_{t=1}(P^{i_{1},j_{t}}_{j}\overline{P}^{j}_{i_{1},j_{t}}+P^{j_{1}\ldots
j_{t-1}j_{t+1},\ldots,j_{3}}_{j}\overline{P}^{j}_{j_{1}\ldots
j_{t-1}j_{t+1},\ldots,j_{3}})\\\nonumber&&=
P^{i_{1},j_{1}}_{j}\overline{P}^{j}_{i_{1},j_{1}}
+P^{j_{2},j_{3}}_{j}\overline{P}^{j}_{j_{2},j_{3}}
+P^{i_{1},j_{2}}_{j}\overline{P}^{j}_{i_{1},j_{2}}\\\nonumber&&+
P^{j_{1},j_{3}}_{j}\overline{P}^{j}_{j_{1},j_{3}}
+P^{i_{1},j_{3}}_{j}\overline{P}^{j}_{i_{1},j_{3}}
+P^{j_{1},j_{2}}_{j}\overline{P}^{j}_{j_{1},j_{2}},
\end{eqnarray}
where $\mathrm{Mat}^{j}_{N_{j},N_{1}N_{2}\cdots
\widehat{N}_{j}\cdots N_{m}}$ is give by matrix (\ref{matrix1}) for
all $N_{j}=2$ and $j=1,2,\ldots, m$. Next, if we assume that the
sequences $I,J$ denote the columns of the
$\mathrm{Mat}^{j}_{N_{j},N_{1}N_{2}\cdots \widehat{N}_{j}\cdots
N_{m}}$, then we can define a class of entanglement monotones for
the multi-qubit states by
\begin{eqnarray}
\mathcal{E}(\mathcal{Q}^{p}_{m}(2,2,\ldots,2))&=&\left(\mathcal{N}\sum^{m}_{j=1,\forall
N_{j}=2} \mathcal{E}_{I,J}(\mathrm{Mat}^{j}_{N_{j},N_{1}N_{2}\cdots
\widehat{N}_{j}\cdots N_{m}}(1,2))\right)^{1/2}.
\end{eqnarray}
This measure for multi-qubit states, which correspond to rows one
and two of matrix (\ref{matrix1}) and its permutations, is invariant
under action of $SL(2,\mathbf{C})$. However, for multi-qubit states
we don't need to used the row indexing $(1,2)$. Now, if we exchange
these rows with any other rows say $\nu,\mu$ of matrix
(\ref{matrix1}), then it is still invariant under action of
$SL(2,\mathbf{C})$. Thus, in this way we can construct an
entanglement measure for general pure multipartite states as follows
\begin{eqnarray}\label{plucker}
\mathcal{E}(\mathcal{Q}^{p}_{m}(N_{1},\ldots,N_{m}))&=&\left(\mathcal{N}_{m}\sum^{m}_{j=1}\sum_{\forall
\nu>\mu=1}\mathcal{E}_{I,J}(\mathrm{Mat}^{j}_{N_{j},N_{1}N_{2}\cdots
\widehat{N}_{j}\cdots N_{m}}(\nu,\mu))\right)^{1/2},
\end{eqnarray}
where $(\mathrm{Mat}^{j}_{N_{j},N_{1}N_{2}\cdots
\widehat{N}_{j}\cdots N_{m}}(\nu,\mu))$ refer to rows $\nu$ and
$\mu$ of matrix (\ref{matrix1}). Now, we can see that entanglement
measure based on the Segre ideals defined in equation (\ref{segre})
coincides with entanglement measure based on the Pl\"{u}cker
coordinate equations of the Grassmann variety defined in equation
(\ref{plucker}). However, for this general case we have shown that
these measure are invariant under repeated  action of
$SL(2,\mathbf{C})$. Thus, for multi-qubit states these measures of
entanglement
 are entanglement monotones but for general pure multipartite
states one need, in general, to show that these measure are
invariant under action of $SL(r,\mathbf{C})$, so these measure of
entanglement should be used with caution for general states. An
alternative way to construct an entanglement monotones for a general
pure multipartite is as follows. Let us consider a quantum system
$\mathcal{Q}^{p}_{m}(N_{1},N_{2},\ldots,N_{m})$, where in this case
we have $r=N_{j}$ and $d=N_{1}N_{2}\cdots \widehat{N}_{j}\cdots
N_{m}$ and let
\begin{eqnarray}
\mathcal{E}_{I,J}(\mathrm{Mat}^{j}_{N_{j},N_{1}N_{2}\cdots
\widehat{N}_{j}\cdots N_{m}})&=&\sum^{r+2}_{t=1}( P^{i_{1},\ldots,
i_{r-1},j_{t}}_{j}\overline{P}^{j}_{i_{1},\ldots,
i_{r-1},j_{t}}\\\nonumber&&+P^{j_{1}\ldots
j_{t-1}j_{t+1},\ldots,j_{r+1}}_{j}\overline{P}^{j}_{j_{1}\ldots
j_{t-1}j_{t+1},\ldots,j_{r+1}}),
\end{eqnarray}
where $I=(i_{1}\ldots i_{r-1}), 1\leq i_{1}<\cdots<i_{r-1}< j_{i}$,
and $J=(j_{1},\ldots ,j_{r+1}), 1\leq j_{1}<\cdots<j_{r+1}\leq d$
are two increasing sequences of numbers from the set
$\{1,2,\ldots,d\}$ and $\mathrm{Mat}^{j}_{N_{j},N_{1}N_{2}\cdots
\widehat{N}_{j}\cdots N_{m}}$ is give by matrix (\ref{matrix1}).
Then we can define  a measure of entanglement for general pure
multipartite states by
\begin{eqnarray}
\mathcal{E}(\mathcal{Q}^{p}_{m}(N_{1},N_{2},\ldots,N_{m}))&=&\left(\mathcal{N}\sum^{m}_{j=1}
\mathcal{E}_{I,J}(\mathrm{Mat}^{j}_{N_{j},N_{1}N_{2}\cdots
\widehat{N}_{j}\cdots N_{m}})\right)^{1/2}.
\end{eqnarray}
Thus, by construction this measure of entanglement is invariant
under action of $SL(r,\mathbf{C})$, that is, a class of entanglement
monotones for multipartite states. However, other properties of this
measure needs further investigation.

\section{Conclusion}
In this paper, we have constructed a class of entanglement monotones
for  multipartite states based on the Segre variety and the
Grassmannian $\mathrm{Gr}(r,d)$, which was  defined in terms of the
Pl\"{u}cker coordinate equations. In particular, we have shown that
entanglement monotones constructed based on Segre variety coincides
with the one constructed by mean of the Pl\"{u}cker coordinate
equations. Thus, we have used some standard but advanced
mathematical tools from complex algebraic geometry to construct a
class of entanglement monotones for multipartite states. However,
the problem of quantifying multipartite entanglement still needs
further investigation and there are many unanswered questions. We
 hope that our result can gives some geometrical insight to
solving such interesting problem of the fundamental quantum theory
with wide application in emerging field of the quantum information
science.
\begin{flushleft}
\textbf{Acknowledgments:}  The  author gratefully acknowledges the
financial support of the Japan Society for the Promotion of Science
(JSPS).
\end{flushleft}


\end{document}